\documentclass[preprint]{aastex631}

\usepackage{CJK}
\usepackage{amsmath}

\usepackage{xcolor}
\received{April 23, 2024}
\revised{August 27, 2024}
\accepted{August 31, 2024}

\submitjournal{ApJ}
\graphicspath{{./}{Figures/}}

\begin{document}

\title{Probing Cold-to-Temperate Exoplanetary Atmospheres: The Role of Water Condensation on Surface Identification with JWST}

\correspondingauthor{Xinting Yu}
\email{xinting.yu@utsa.edu}

\author[0000-0002-7479-1437]{Ziyu Huang\begin{CJK*}{UTF8}{gbsn}
(黄子钰)\end{CJK*}}
\affiliation{Center for Space Physics, 
Boston University\\
685 Commonwealth Ave\\
Boston, MA 02215, USA}

\author[0000-0002-7479-1437]{Xinting Yu\begin{CJK*}{UTF8}{gbsn}
(余馨婷)\end{CJK*}}
\affiliation{Department of Physics and Astronomy, University of Texas at San Antonio\\
1 UTSA Circle\\
San Antonio, TX 78249, USA}

\author[0000-0002-8163-4608]{Shang-Min Tsai}
\affiliation{Department of Earth and Planetary Sciences, University of California Riverside\\
900 University Ave \\
Riverside, CA 92521, USA}

\author[0000-0002-8837-0035]{Julianne I. Moses}
\affiliation{Space Science Institute, \\
Boulder, Colorado 80301, USA}

\author[0000-0003-3290-6758]{Kazumasa Ohno}
\affiliation{National Astronomical Observatory of Japan\\
2 Chome-21-1 Osawa \\
Mitaka City, Tokyo 181-8588, Japan}

\author[0000-0001-6878-4866]{Joshua Krissansen-Totton}
\affiliation{Department of Earth and Space Sciences\\
University of Washington, Seattle\\
1410 NE Campus Pkwy \\
Seattle, WA 98195, USA}

\author[0000-0002-8706-6963]{Xi Zhang}
\affiliation{Department of Earth and Planetary Sciences, University of California Santa Cruz\\
1156 High St\\
Santa Cruz, CA 95064, USA}

\author[0000-0002-9843-4354]{Jonathan J. Fortney}
\affiliation{Department of Astronomy and Astrophysics \\
University of California Santa Cruz\\
1156 High Street\\
Santa Cruz, California 95064, USA}

\begin{abstract}
Understanding the surface temperature and interior structure of cold-to-temperate sub-Neptunes is critical for assessing their habitability, yet direct observations are challenging. In this study, we investigate the impact of water condensation on the atmospheric compositions of sub-Neptunes, focusing on the implications for JWST spectroscopic observations. By modeling the atmospheric photochemistry of two canonical sub-Neptunes, K2-18 b and LHS 1140 b, both with and without water condensation and with and without thick atmospheres, we demonstrate that water condensation can significantly affect the predicted atmospheric compositions. This effect is driven by oxygen depletion from the condensation of water vapor and primarily manifests as an increase in the C/O ratio within the photochemically active regions of the atmosphere. This change in composition particularly affects planets with thin H$_2$-dominated atmospheres, leading to a transition in dominant nitrogen and carbon carriers from N$_2$ and oxygen-rich species like CO/CO$_2$ towards heavier hydrocarbons and nitriles. While our models do not fully account for the loss mechanisms of these higher-order species, such molecules can go on to form more refractory molecules or hazes. Planets with thin H$_2$-rich atmospheres undergoing significant water condensation are thus likely to exhibit very hazy atmospheres. The relatively flat JWST spectra observed for LHS 1140 b could be consistent with such a scenario, suggesting a shallow surface with extensive water condensation or a high atmospheric C/O ratio. Conversely, the JWST observations of K2-18 b are better aligned with a volatile-rich mini-Neptune with a thick atmosphere.
\end{abstract}

\keywords{Exoplanet atmospheres (487); Exoplanet atmospheric composition (2021);
Exoplanet surfaces (2118); Extrasolar gaseous planets (2172); Extrasolar rocky planets (511)}

\section{Introduction} \label{sec:intro}

The James Webb Space Telescope (JWST) has inaugurated a new era in astronomy, particularly in enhancing our understanding of exoplanets, including their evolution, atmospheric composition, and habitability. However, the nature of mid-sized exoplanets, especially those within the 1.7 to 3.5 Earth radii range, or the so-called ``sub-Neptunes" \citep{2017AJ....154..109F}, remains unclear, as a range of interior compositions can explain a planet's observed mass and radius. There are ongoing debates about the classification of sub-Neptunes as either super-Earths (similar to rocky terrestrial bodies with thin atmospheres over solid surfaces) or gas-rich mini-Neptunes (similar to gas giants with thick atmospheres and deep, hot, or perhaps no well-defined surfaces) \citep{yu2021identify,2021JGRE..12606639B}. Some of the super-Earths may be the so-called ``Hycean" worlds, which have H$_2$ atmospheres with ocean-covered surfaces \citep{madhusudhan2023chemical}. Understanding conditions at the potential atmosphere-surface interface on sub-Neptunes, such as surface temperature and pressure, is crucial for determining their nature and classification. While direct observation of surface conditions is challenging, atmospheric characterization offers a promising approach to breaking the interior composition degeneracies and indirectly inferring the lower boundary conditions of sub-Neptunes \citep{yu2021identify, hu2021unveiling, tsai2021inferring}.

In previous work, \cite{yu2021identify} proposed that the study of atmospheric chemistry could reveal conditions at the core-atmosphere boundary for sub-Neptunes with hydrogen-dominated atmospheres. In cases where the atmosphere is relatively thin (for instance, less than 10 bars), gases that are destroyed by photochemistry in the upper atmosphere, such as ammonia (NH$_3$) and methane (CH$_4$), tend to diminish with time as there is no deeper, hotter part of the atmosphere where these molecules could be replenished by thermochemical reactions. \cite{hu2021unveiling} examined the unique characteristics of an ocean planet enveloped by a thin H$_2$-rich atmosphere, where a balance of gas solubility is anticipated. A canonical sub-Neptune planet K2-18 b, which has recently been the target of space-based spectroscopic observations \citep{benneke2017spitzer,tsiaras2019water, madhusudhan2023carbon}, was used as a reference in both pieces of research. While \citet{yu2021identify} and similar work by \citet{tsai2021inferring} did not account for the exchange of gases between the atmosphere and the surface, \cite{hu2021unveiling} and \cite{madhusudhan2023chemical} considered a range of CO$_2$ levels that would be compatible with a global habitable liquid water ocean. Despite these differences, all the above studies agreed that the lack of NH$_3$ is the most responsive indicator for detecting shallow surfaces. In addition, the abundance of CH$_4$ in a 1-bar atmosphere with temperate surface temperatures would be significantly lower than in a deep atmosphere case, regardless of CO$_2$ levels, unless there is a large surface flux of methane mimicking a methane-generating biosphere \citep{wogan2024jwst}. 

Recently, the transmission spectra of two canonical temperate sub-Neptunes in the habitable zone, obtained by JWST, have become available, including K2-18 b \citep{madhusudhan2023carbon} and LHS-1140 b \citep{2024ApJ...968L..22D, 2024arXiv240312617D, 2024arXiv240615136C}. The absence of H$_2$O and NH$_3$ and enrichment in CO$_2$ and CH$_4$ led to the suggestion that K2-18 b could be a possible ``hycean" world with a liquid water ocean underneath a thin H$_2$-dominated atmosphere cold enough for H$_2$O ice to condense \citep{madhusudhan2023carbon} or a gas-rich mini-Neptune with water-condensation below the observable photosphere \citep{wogan2024jwst}. No matter the nature of the planet, the low water abundance in K2-18 b's atmosphere is likely a result of water condensation. LHS-1140 b is found to have a rather flat spectrum \citep{2024ApJ...968L..22D, 2024arXiv240312617D, 2024arXiv240615136C}. The colder zero-albedo equilibrium temperature (234 K) of LHS 1140 b versus K2-18 b (278 K) suggests water condensation would likely occur in its atmosphere as well.

Water condensation is not only crucial for a planet to hold onto its water and affects the planet's habitability, but it can also impact the atmospheric chemistry of sub-Neptunes \citep{2021ApJ...921...27H,madhusudhan2023chemical}. However, a systematic exploration of the role of water condensation and its sensitivity to model lower boundary conditions is still lacking. In this study, following the framework proposed by \cite{yu2021identify}, we aim to provide an extended surface-identification framework for cold-to-temperate exoplanets where water condensation is important. The findings from our study can help us better understand the existing and upcoming JWST spectroscopic observations of cold/temperate sub-Neptunes, potentially breaking their interior composition degeneracies. In addition, exploring the surface conditions of cold/temperate sub-Neptunes is a crucial step in searching for and characterizing potentially habitable worlds, particularly in discerning the proportion of super-Earth-like habitable exoplanets within the sub-Neptune population. Gaining a thorough understanding of the nature of these potentially habitable sub-Neptunes can also help guide the design of future exoplanet missions. 

\section{Methods} \label{sec:style}
\subsection{Modeling Target}

While our modeling is designed to represent generic cold-to-temperate sub-Neptunes with water condensation, we adopt the parameters of K2-18 b and LHS 1140 b for this study. We adopt K2-18 b's mass, orbital radius, and stellar radius from \citet{2019A&A...621A..49C} and planet radius from \citet{benneke2019water}. The mass and radius of LHS 1140 b are adopted from \citet{2024ApJ...960L...3C}, and its orbital radius and stellar radius are from \citet{2019AJ....157...32M}. The atmosphere of K2-18 b has been characterized by the Kepler Space Telescope (K2) in the $0.4-0.9~\mu$m range, the Hubble Space Telescope (HST) Wide Field Camera 3 (WFC3) in the $1.1-1.7~\mu$m range, and the Spitzer Telescope at 3.6 and $4.5~\mu$m \citep{benneke2017spitzer,benneke2019sub,tsiaras2019water}, and JWST from $0.6-5~\mu$m \citep{madhusudhan2023carbon}. The atmosphere of LHS 1140 b has been characterized by the Hubble Space Telescope (HST) Wide Field Camera 3 (WFC3) in the $1.1-1.7~\mu$m range \citep{edwards2020hubble} and JWST from $0.6-2.8~\mu$m \citep{2024arXiv240312617D, 2024arXiv240615136C} and $1.67-5.18~\mu$m  \citep{2024ApJ...968L..22D}.

\subsection{Atmospheric Structure}
 
For a mini-Neptune-like exoplanet with a thick atmosphere, creating a pressure-temperature (P-T) profile within our model means accounting for both incident radiation from the parent star and thermal emission from the planetary atmosphere and interior \citep{2005ApJ...627L..69F}. We employ a well-established one-dimensional radiative-convective equilibrium model \citep{marley1999thermal, Marley&Robinson15} to compute the atmospheric structure. %The Python version of the adopted model is now publicly available \citep{Mukherjee+23_PICASO}. 
We adopt the k-coefficient opacity table from \citet{Lupu+22}. Since the opacity data is not available at $<{10}^{-6}~{\rm bar}$, we fixed the opacity at these lower pressures to that at $P={10}^{-6}~{\rm bar}$.

Here, we consider two sub-Neptune planets that are cold enough for water to condense, significantly reducing the expected H$_2$O mixing ratio above the tropopause cold trap, thereby affecting stratospheric chemistry: K2-18 b and LHS 1140 b. The zero-albedo equilibrium temperature is 278 K for K2-18 b and 234 K for LHS 1140 b. We generate P-T profiles of both planets with a grid of varying semi-major axis values to achieve the bolometric stellar flux equivalent to the atmosphere with desired bond albedos (A$\rm_b$) of 0, 0.3, 0.56, and 0.75. The choice of bond albedo of 0.3 is reasonable because Earth, Mars, Titan, and the giant planets in the Solar System all have bond albedos around 0.2-0.4 \citep{2001plsc.book.....D}, as well as some hot-Jupiters \citep{2005ApJ...626..523C,2021NatAs...5.1001H}. The higher albedo cases represent the highly reflective clouds/hazes that can prevent runaway greenhouse \citep{innes2023runaway, 2023ApJ...944...20P}. By considering a range of bond albedos for both planets, we can investigate how variations in water condensation influence the atmospheric chemistry of sub-Neptunes with different surface conditions. Overall, the range of albedos leads to an equilibrium temperature for K2-18 b of 278 K (A$\rm_b=0$), 254 K (A$\rm_b=0.3$), 227 K (A$\rm_b=0.56$), and 197 K (A$\rm_b=0.75$) and for LHS 1140 b of 234 K (A$\rm_b=0$), 214 K (A$\rm_b=0.3$), 191 K (A$\rm_b=0.56$), and 165 K (A$\rm_b=0.75$).

\begin{figure}[h!]
    \centering
        \includegraphics[width = \textwidth]{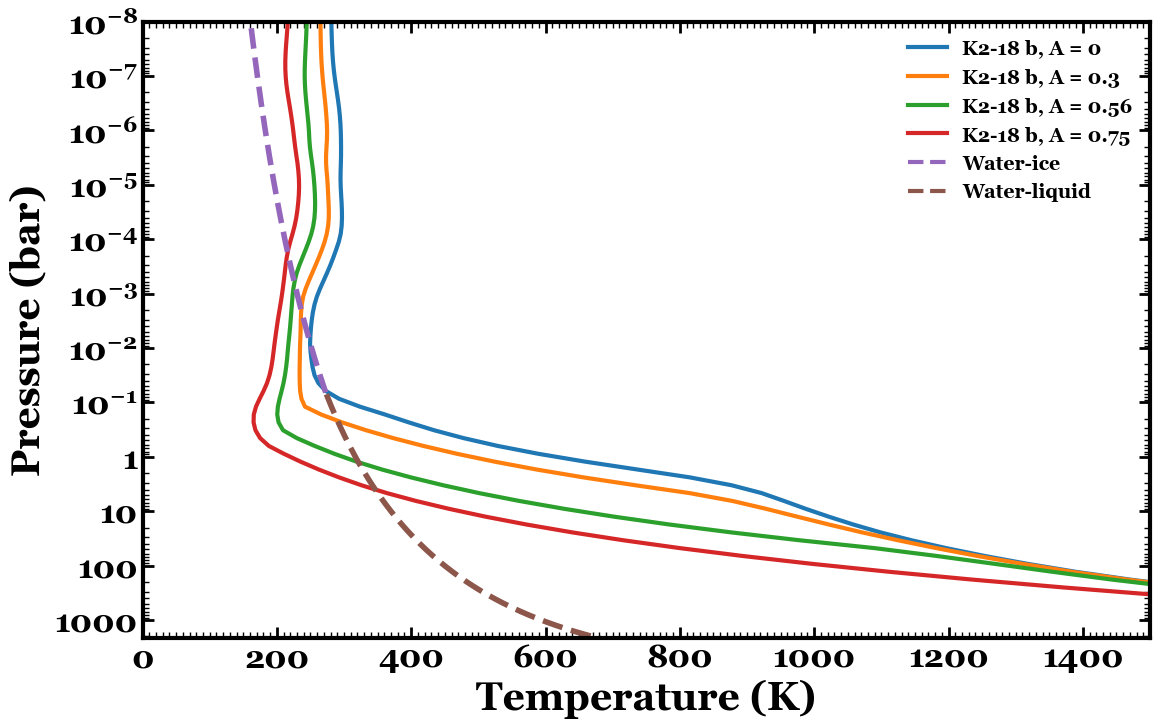}  
        \includegraphics[width = \textwidth]{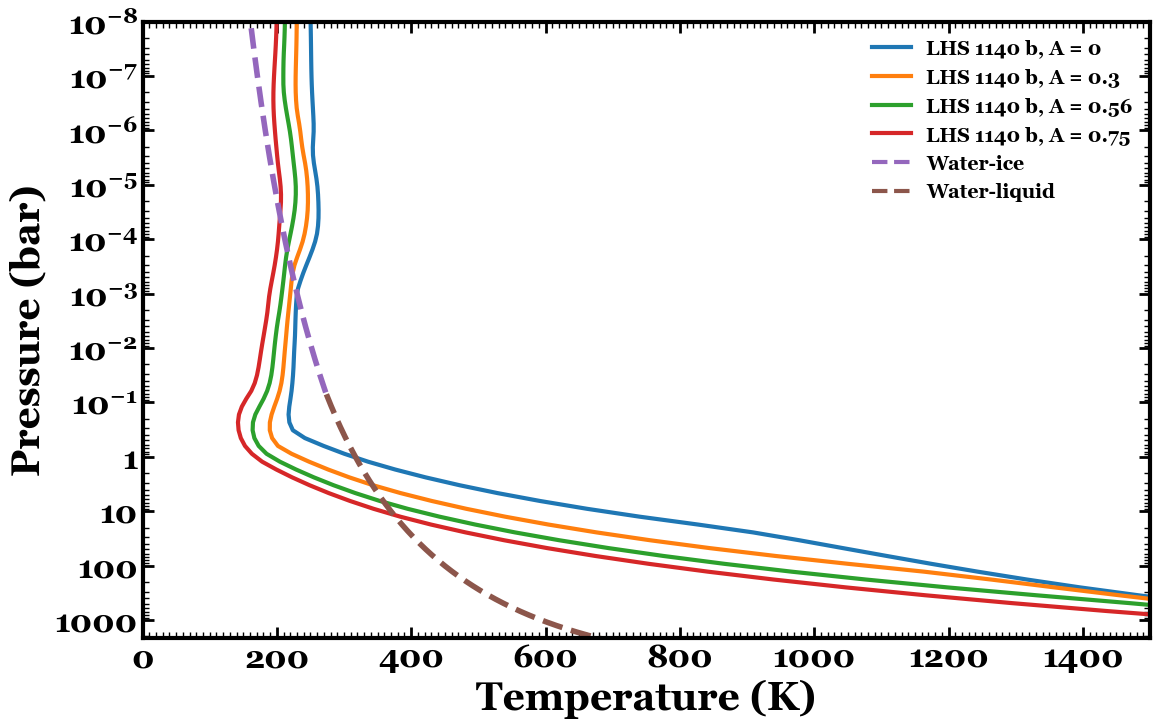} 
    \caption{Pressure temperature profiles of K2-18 b (top) and LHS 1140 b (bottom) with albedo from 0 to 0.75. Dashed lines are condensation curves of H$_2$O (purple for sublimation and brown for vaporization) following Eq.~\ref{eq:conden_ice} and Eq.~\ref{eq:conden_liquid}. These condensation curves are constructed by using e$_s$/X$_{H2O}$, where we assumed the mixing ratio of water, X$_{H2O}=0.1$.}
\label{fig:PTprofile}    
\end{figure}

For both planets, we assume a 100 times solar metallicity before accounting for water condensation, which is plausible from formation models for sub-Neptunes \citep{2013ApJ...775...80F, 2016ApJ...831...64T} and is also consistent with the inferred metallicity for carbon from JWST observations of K2-18 b \citep{madhusudhan2023carbon}. We assume an internal heat flux of $T_ {\rm int} = 70~K$, for consistency with \citet{yu2021identify}. Given these metallicity and temperature-profile assumptions, water starts to condense between $0.05-10~\rm bar$ for both planets. The input stellar spectra adopted for the radiative-convective equilibrium models are interpolated from the Phoenix stellar spectra grid through \texttt{pysynphot} package \citep{STScI}. The generated P-T profiles are shown in Fig.~\ref{fig:PTprofile}. We indicate the condensation curves of water as dashed lines (purple for the gas-solid transition and brown for the gas-liquid transition). A previous study by \cite{tsai2021inferring} has shown that surface pressure has minimal effects on the atmospheric thermal structure. As a result, we cut off the P-T profile with a surface pressure of 1 bar to investigate the case of exoplanets with thin, shallow atmospheres above a solid surface (hereafter simply called the shallow-surface cases).

The saturation vapor pressure of H$_2$O for ice state and liquid state is calculated from Eq. A2a and Eq. A2b from \cite{ackerman2001precipitating}:
\begin{equation}
e_s\left(\mathrm{H}_2 \mathrm{O}, \text { ice}\right)=6111.5 \exp \left(\frac{23.036 T_{\mathrm{C}}-T_{\mathrm{C}}^2 / 333.7}{T_{\mathrm{C}}+279.82}\right)
\label{eq:conden_ice}
\end{equation}

\begin{equation}
e_s\left(\mathrm{H}_2 \mathrm{O}, \text { liquid}\right)=6112.1 \exp \left(\frac{18.729 T_{\mathrm{C}}-T_{\mathrm{C}}^2 / 227.3}{T_{\mathrm{C}}+257.87}\right)
\label{eq:conden_liquid}
\end{equation}
with $e_s$ in $\rm dynes~\rm cm^{-2}$ and the atmospheric temperature $T_{\rm C}$ in degree Celsius.

\subsection{1-D Photochemical Model}
 
In this study, we use the 1-D photochemical modeling tool VULCAN \citep{tsai2017vulcan,tsai2021comparative} to model the temporal evolution of the chemical composition of both planets. The following equation is solved for the balance between chemical reactions and vertical mixing:

\begin{equation}
	\frac{\partial n_i}{\partial t}=\mathcal{P}_i-\mathcal{L}_i-\frac{\partial \phi_i}{\partial z},
\end{equation}

where $n_i$ is the number density of species $i$, $\mathcal{P}_i$ and $\mathcal{L}_i$  are the chemical production and loss rates for each species, and $\frac{\partial \phi_i}{\partial z}$ is the gradient of the vertical transport flux of species $i$; the flux is assumed to be controlled by eddy and molecular diffusion. In this study, we follow \cite{tsai2021inferring} by using the following equation to describe the eddy diffusion coefficient $K_{zz}$ for pressure $P$ $<$ 1 bar and assumed constant $K_{\rm zz} = 10^5~{\rm cm^2~s^{-1}}$ for $P$ $\ge$ 1 bar:

\begin{equation}
K_{\mathrm{zz}}=10^5\left(\frac{1 \ \mathrm{bar}}{P}\right)^{0.4}~\mathrm{~cm}^{2}~\mathrm{s}^{-1}
\label{eq:Kzz}
\end{equation}

%Note that at the temperature minimum for both of our model planets, sulfur is expected to condense out and therefore will not be present in appreciable quantities in the stratosphere; thus, we did not include sulfur in our chemical network.

The details of the model, including its numerical scheme, can be found in \cite{tsai2021comparative}. The chemical reaction network includes C, N, O, H, and a total of 57 species with 448 forward chemical reactions, including both thermochemistry and photochemistry. Similarly to \cite{yu2021identify}, we assume a hydrogen-dominated and metal-rich atmosphere of 100 times solar metallicity. We adopt the stellar spectrum of an active M2 star at the age of 45 Myr from HAZMAT \citep{2020ApJ...895....5P} as in \cite{tsai2021inferring} for K2-18 b and the stellar spectrum of Proxima Centauri (GJ 551) from the MUSCLES treasury survey \citep{2016ApJ...820...89F, 2017ApJ...843...31Y} as in \citet{2021A&A...647A..48W} for LHS 1140 b. We used the same bolometric UV flux regardless of the assumed bond albedo for each planet. For the upper and lower boundaries of our atmospheric models, we adopt zero flux conditions for all species consistent with the approach taken by \citet{yu2021identify}. This assumption implies the absence of sources and sinks at the surface, as well as the absence of atmospheric escape or influx at the upper boundary. The expected mass loss for both K2-18b and LHS 1140 b due to atmospheric escape --- less than 1\% of the mass of the planet over the age of the system amounts as estimated by \citet{2020A&A...634L...4D} and \citet{2023MNRAS.525.5168M} --- supports the validity of our zero upper boundary flux assumption. While geological and biological activities could potentially introduce non-zero fluxes at the lower boundary, particularly for ``super-Earth"-like exoplanets, the specific rates of such processes under exoplanetary conditions remain largely unknown. In light of these uncertainties, our zero-flux boundary conditions serve as a zero-order model to estimate the influence of the surface on atmospheric composition.

For water condensation, we convert the gas phase water to the condensed phase when the partial pressure of gas phase H$_2$O exceeds the saturation vapor pressure. Because the timescale of the condensation is much shorter compared to chemical evolution, here we calculate the condensation in the first time step. Within the saturation region, which is the region from the bottom level of condensation to the pressure level ($P_0$) where the water saturation mixing ratio is the minimum, we set the mixing ratio of water to its saturated value. Beyond this pressure level, we let the water mixing ratio freely evolve in the upper atmosphere ($P<P_0$). For the different albedo cases of both K2-18 b and LHS 1140 b, $P_0$ is between 1 and 10$^{-2}$ bar. The photochemical model evolves to a steady state using the same convergence criteria as \citet[Equation 11]{tsai2017vulcan}. Because of photochemistry and water condensation, which removes gas-phase water into the condensed phase, the final atmospheric molecular composition can differ significantly from the initially assumed 100 times solar composition, despite the conservation of mass/elements. In this study, we consider two representative scenarios for the atmospheric depth, namely the shallow-surface (surface pressure of $1~{\rm bar}$) and the deep-surface (surface pressure of $1000~{\rm bar}$) scenarios. 

\subsection{Transmission Spectra Modeling}

We utilize an open-source radiative transfer code petitRADTRANS \citep{molliere2019petitradtrans} to generate transmission spectra of our modeling targets. The atmospheric opacities in these models use the correlated-$k$ opacity tabulation method. The generation of the $k$-tables adheres strictly to the methodologies established by \citet{molliere2019petitradtrans}. We include opacity data of major molecules: CO, CO$_2$, H$_2$O, CH$_4$, NH$_3$, HCN, and the continuum opacity due to H$_2$-H$_2$ and H$_2$-He collision-induced absorption.

\section{Results and Discussion}
\subsection{The Effect of Surface Pressure on the Atmospheric Composition of Cold to Temperate Sub-Neptunes with Water Condensation}

\begin{figure}
    \centering
        \includegraphics[width = \textwidth]{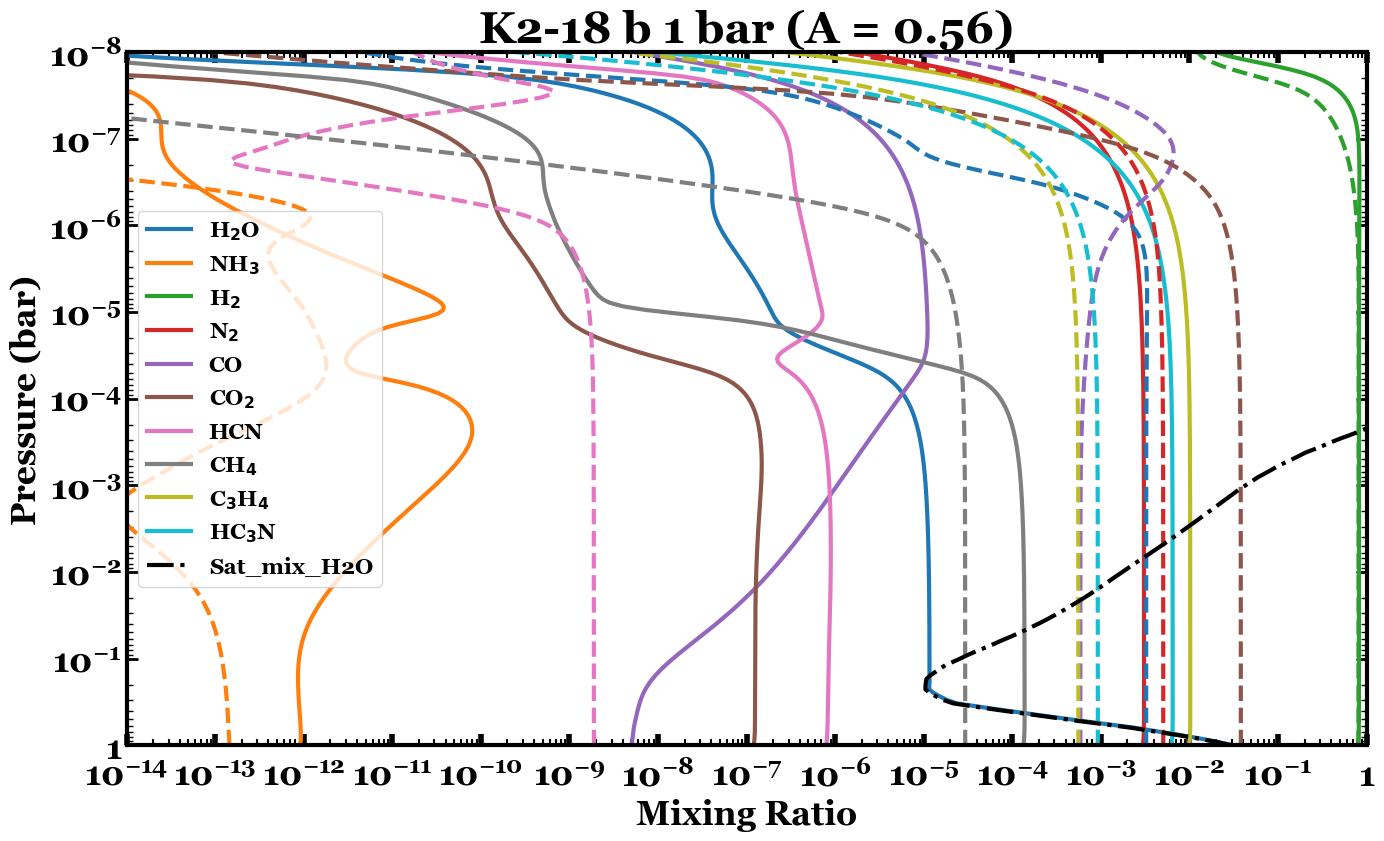}  
        \includegraphics[width = \textwidth]{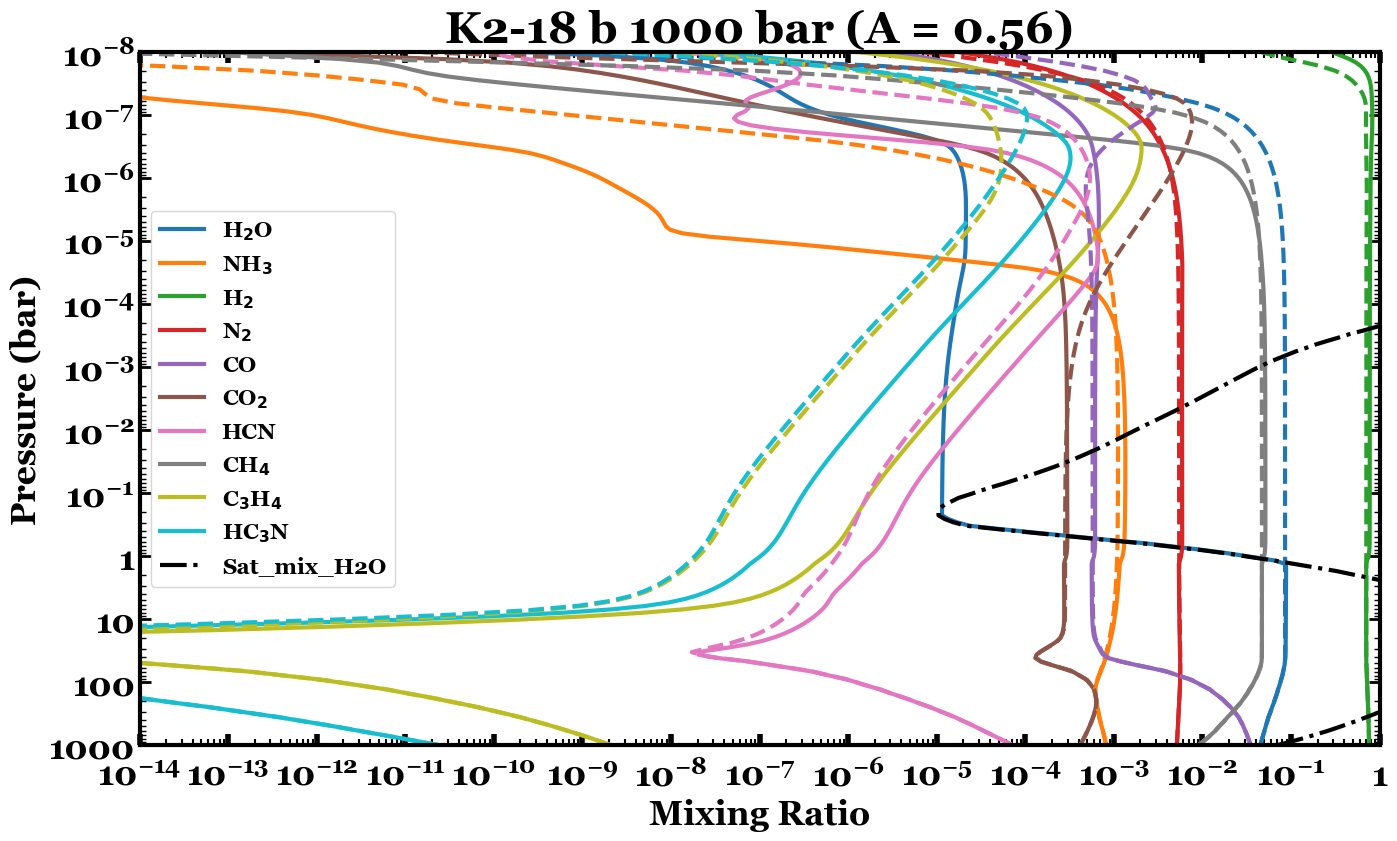} 
    \caption{Simulated volume mixing ratio (VMR) profiles for the main chemical species for K2-18 b with A$\rm_b=0.56$ (T$_{\rm eq} = 227~\rm K$) for the shallow case and deep case, with and without water condensation (solid and dashed lines), the black dash-dotted line shows the saturation vapor mixing ratio of water.}
\label{fig:K2_condense}    
\end{figure}

\begin{figure}[h!]
    \centering
        \includegraphics[width = \textwidth]{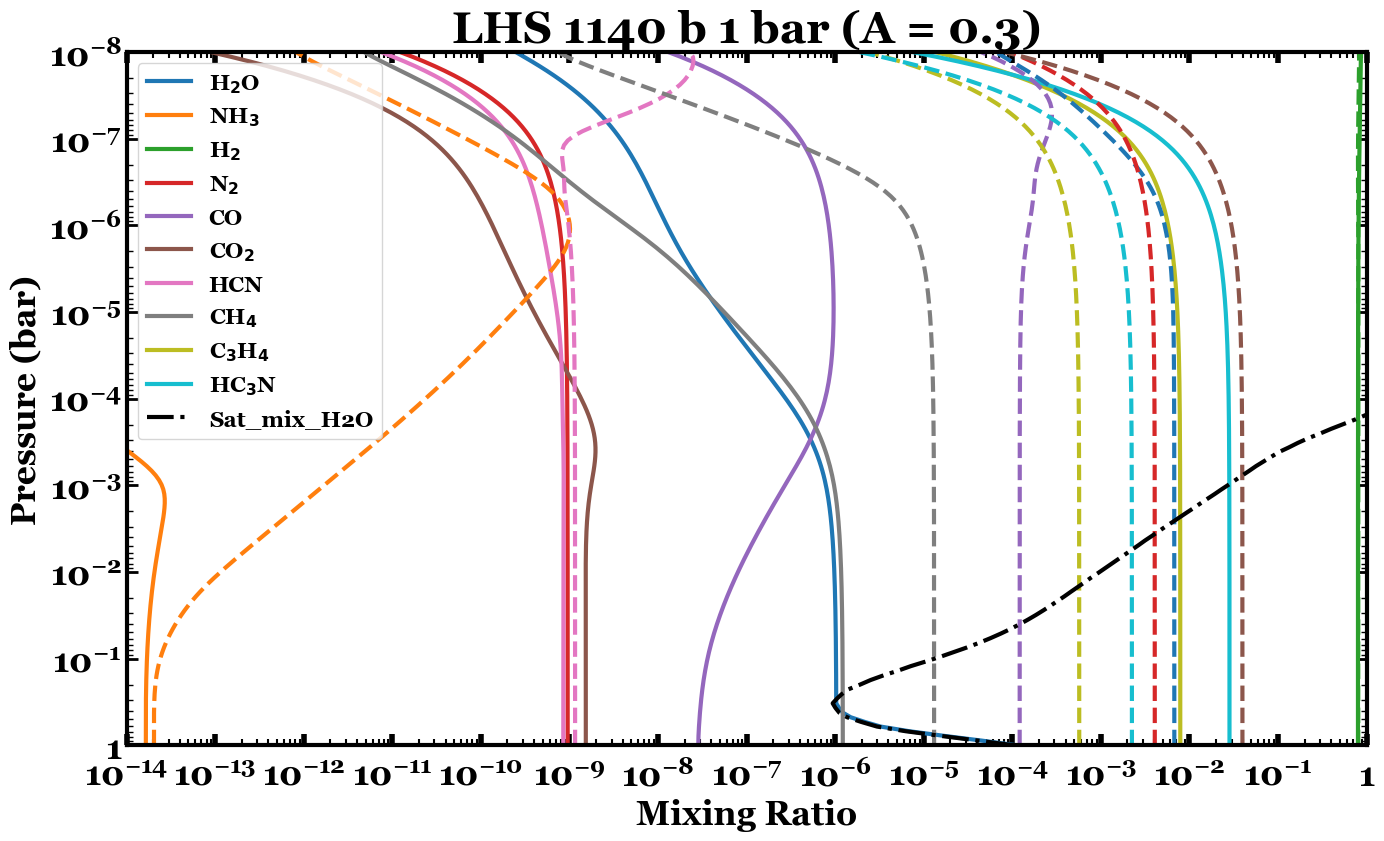} 
        \includegraphics[width = \textwidth]{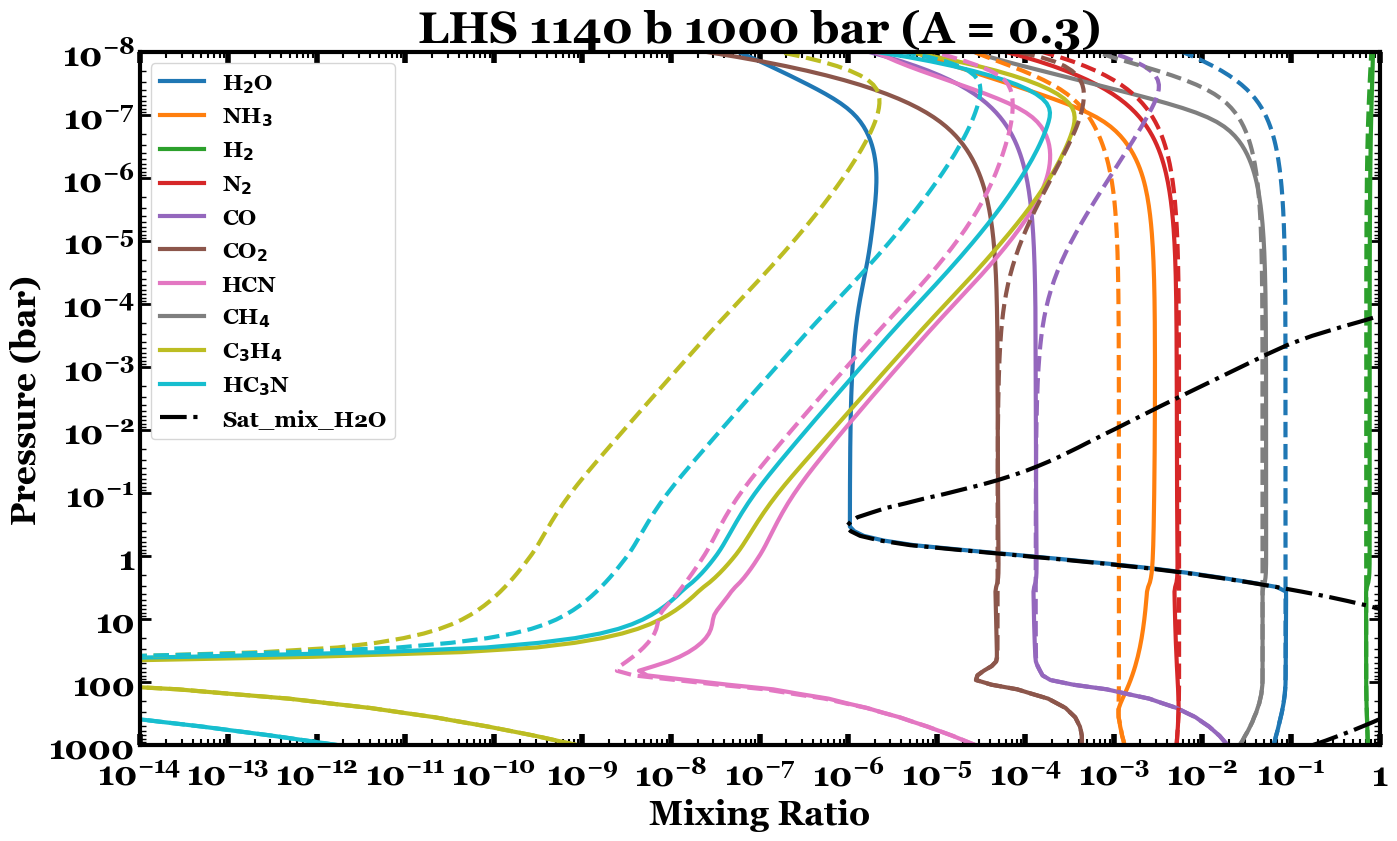} 
    \caption{Simulated VMR profiles for the main chemical species for LHS 1140 b with A$\rm_b=0.3$ (T$_{\rm eq} = 214~\rm K$) for the shallow and deep cases, with and without water condensation. The notations are the same as in Figure \ref{fig:K2_condense}.}
\label{fig:LHS_condense}    
\end{figure}

We first examine the effect of water condensation on the final, steady-state atmospheric composition on a sub-Neptune with a shallow surface and a deep surface. Here, we select the albedo cases of A$\rm_b=0.56$ for K2-18 b and A$\rm_b=0.3$ for LHS 1140 b, comparing scenarios with and without condensation (Figures~\ref{fig:K2_condense} and \ref{fig:LHS_condense}). The steady-state atmospheric compositions for other albedo cases for K2-18 b and LHS 1140 b can be found in the Appendix (Figures~\ref{fig:K2_appendix} and \ref{fig:LHS_appendix}). We choose to present the A$\rm_b=0.56$ case for K2-18 b in the main text because, at this albedo, water is removed by condensation sufficiently to match the retrieved water mixing ratio from JWST \citep{madhusudhan2023carbon}. For LHS 1140 b, albedo cases of A$\rm_b=0.3$ and above can all match the JWST measured flat spectra \citep{2024arXiv240312617D, 2024arXiv240615136C, 2024ApJ...968L..22D}, see also Section~\ref{sec:trans}.

For the deep-surface scenarios of both K2-18 b and LHS 1140 b albedo cases presented (Figures~\ref{fig:K2_condense} and \ref{fig:LHS_condense}), water condensation does not affect the water mixing ratio in the deep atmosphere, where temperatures are high enough to prevent condensation. As a result, the mixing ratios of key species such as CH$_4$, N$_2$, NH$_3$, CO, and CO$_2$ are governed by thermochemical equilibrium in the deep atmosphere and then quenched at around 10-100 bar. In the upper atmosphere, however, water condensation plays a significant role in shaping the mixing ratios of CO and CO$_2$ for K2-18 b. Without water condensation, H$_2$O photolysis leads to the formation of OH radicals that effectively convert CO into CO$_2$. Consequently, the mixing ratios of CO and CO$_2$ are altered, with CO$_2$/CO $>$ 1 starting from $P < 10^{-4}$ bar. However, when we include water condensation, there is a noticeable increase in CO relative to CO$_2$ above the cloud level, as CO$_2$/CO $<$ 1 holds true for all pressure levels from $P > 1\rm$ bar. This change is due to the limited availability of OH (and H) from H$_2$O photolysis in the condensation case, which makes CO less efficiently converted to CO$_2$, and CH$_4$ less efficiently converted to CO, see discussions in \cite{1973JAtS...30.1437M, 1994Icar..111..124N, 2016ApJ...824..137Z, 2016ApJ...829...66M, hu2021photochemistry, tsai2021comparative}. For LHS 1140 b, however, CO$_2$/CO $<$ 1 holds true for the deep case both with and without condensation. This is due to the lower stellar UV flux used for LHS 1140 b compared to K2-18 b, which leads to less effective formation of OH from H$_2$O photolysis, and thus the CO$_2$/CO ratio is always less than 1. Overall, the abundance of key species for the deep-surface scenarios with water condensation, except for CO, CO$_2$, C$_3$H$_4$, and HC$_3$N, closely mirrors the scenarios without water condensation in the upper atmosphere.

For the shallow-surface case, if the planet is warm enough that water does not condense, the shallow-surface case results in lower CH$_4$ and higher CO$_2$ and CO mixing ratios compared to the deep-surface case, as a result of insufficient thermochemical recycling of CH$_4$ \citep[e.g.,][]{yu2021identify}. H$_2$O and CH$_4$ together are converted photochemically to more stable species such as CO$_2$ and CO, as described in \cite{yu2021identify}, see results in dashed lines for K2-18 b and LHS 1140 b in Figures~\ref{fig:K2_condense} and \ref{fig:LHS_condense}, when we turn water condensation off. 

When we include water condensation, however, there is a marked departure from the non-condensation scenario due to the insufficient water supply from the deep atmosphere; see the results in solid lines in Figures~\ref{fig:K2_condense} and \ref{fig:LHS_condense}. For both K2-18 b and LHS 1140 b, because water vapor is depleted both within the tropospheric water saturation region and above the tropopause cold trap, the amount of oxygen within the system is reduced, circumventing the efficient conversion of CH$_4$ to CO and CO$_2$ through photochemical processes. Instead, the CH$_4$ is preferentially converted to higher-order hydrocarbons and nitriles such as HC$_3$N and C$_3$H$_4$, which do not contain oxygen. For LHS 1140 b, water is being removed above the cloud with a mixing ratio of $1\times10^{-6}$, and the mixing ratio of CO$_2$ and CO both drop below 1 ppm, much lower than the mixing ratios for the non-condensation shallow surface case. The increasing dominance of hydrocarbons and nitriles is a direct result of the high C/O ratio in the atmospheric region above the tropopause cold trap, where photochemistry dominates, the lack of thermochemical recycling back to CH$_4$ and NH$_3$ when the atmosphere does not extend deep, and the assumption that the hydrocarbons/nitriles are not lost to the surface. 

Note that the formation of high-order organics, such as HC$_3$N and C$_3$H$_4$, are a result of the limited chemical network, which does not contain sufficient destruction reactions for these species. One should, therefore, not consider the results for these particular species as realistic. Instead, these abundant species can be treated as proxies for the production of organic-rich heavy molecules and hazes that much of the carbon and nitrogen will be transferred into via photochemistry. Ultimately, the more refractory species will condense and rain out, removing most of that amount of carbon and nitrogen from the atmosphere if there are no surface/interior processes that recycle the species back to CH$_4$ and NH$_3$ or N$_2$. During the transition phase from CH$_4$ and NH$_3$ to these photochemically produced species, the atmosphere will most likely be very hazy, similar to Titan, as is discussed in \citet{yu2021identify} and \citet{madhusudhan2023chemical}. A more detailed discussion of the formation of high-order organics and hazes can be found in Section~\ref{sec:heavy}.

With the condensation of water, the ammonia mixing ratio remains similar to the case without condensation. NH$_3$ is depleted to VMRs of $10^{-12}$ for K2-18 b and $10^{-14}$ for LHS 1140 b with shallow surfaces. However, for both planets in the shallow-surface cases, nitrogen speciation changes as a result of the depletion of water vapor in the system. In addition to molecular nitrogen, N is also deposited in heavier-order nitriles such as HC$_3$N. The major reactions leading to HC$_3$N are listed as:

\[\begin{aligned} & \mathrm{HCN}+h \nu \rightarrow \mathrm{CN}+\mathrm{H}, \\ & \mathrm{CH}_4+\mathrm{H} \rightarrow \mathrm{CH}_3+\mathrm{H}_2, \\ & \mathrm{CH}_3+\mathrm{CH}_3+\mathrm{M} \rightarrow \mathrm{C}_2 \mathrm{H}_6+\mathrm{M}, \\ & \mathrm{C}_2 \mathrm{H}_6+\mathrm{H} \rightarrow \mathrm{C}_2 \mathrm{H}_5+\mathrm{H}_2, \\ & \mathrm{C}_2 \mathrm{H}_5+\mathrm{H} \rightarrow \mathrm{C}_2 \mathrm{H}_4+\mathrm{H}_2, \\ & \mathrm{C}_2 \mathrm{H}_4+h \nu \rightarrow \mathrm{C}_2 \mathrm{H}_2+\mathrm{H}+\mathrm{H}, \\ & \mathrm{CN}+\mathrm{C}_2 \mathrm{H}_2 \rightarrow \mathrm{HC}_3 \mathrm{N}+\mathrm{H},\end{aligned}\]

Overall, the impact of water condensation on atmospheric photochemistry varies significantly depending on the surface conditions of the planets. For planets with a deep surface, water condensation minimally affects the compositions of primary atmospheric species, as the mixing ratios of major atmospheric species like CH$_4$, N$_2$, NH$_3$, CO, and CO$_2$ are governed by thermochemical equilibrium in the deep atmosphere and transport-induced quenching. Conversely, planets with a shallow surface experience more pronounced changes due to water condensation. 

If the water saturation vapor pressure at the planet's surface is comparable to or greater than the atmospheric H$_2$O partial pressure, as demonstrated by the warmer K2-18 b cases with albedos less than 0.56 as shown in the Appendix Figure~\ref{fig:K2_appendix}, the atmospheric composition for the shallow-surface case still has abundant CH$_4$, CO, and CO$_2$. However, as the atmosphere gets cooler, which causes more water condensation, heavier-order hydrocarbons and nitriles begin to accumulate in the atmosphere, which would result in increased haziness compared to the shallow-surface case without water condensation.

In the case where the saturation vapor pressure at the surface is substantially lower compared to the partial pressure of water, as demonstrated in the K2-18 b and LHS 1140 b albedo case presented in Figures~\ref{fig:K2_condense} and \ref{fig:LHS_condense}, the saturation vapor mixing ratio restricts the total availability of atmospheric oxygen. The depletion of the original nitrogen and carbon carriers, notably NH$_3$ and CH$_4$, still occurs due to the presence of a shallow surface. This is because the temperature of the shallow surface is insufficient for thermochemistry to recycle photochemically produced hydrocarbon and nitrogen-bearing species, similar to the scenario without water condensation as described by \citet{yu2021identify}. However, due to the lack of oxygen in the system caused by water condensation, the C/O ratio in the gas-phase atmosphere increases compared to the case with no water condensation. As a result, the main carbon and nitrogen carriers (NH$_3$ and CH$_4$) are being taken over by oxygen-poor species for the case with water condensation, including higher-order hydrocarbons and nitriles, along with molecular nitrogen, instead of being converted to oxygen-rich species such as CO, CO$_2$, and molecular nitrogen for the case without water condensation.

Thus, for a sub-Neptune with an equilibrium temperature that leads to water condensation in its atmosphere, the depletion of CO$_2$, CO, CH$_4$, and NH$_3$, along with likely enhanced haziness due to the formation of refractory hydrocarbons and nitriles, may indicate a shallow surface and significant water depletion near the surface through water condensation. This condition suggests insufficient oxygen in the atmosphere to maintain higher levels of these molecules, leading to the accumulation of oxygen-poor hydrocarbons and nitriles. Such a type of water-depleted, photochemically evolved exoplanet is, in principle, very similar to the atmospheric evolution of Titan, which results in an N$_2$-dominated atmosphere with abundant higher-order hydrocarbons and nitriles. 

Our shallow-surface exoplanet cases with water condensation differ from Titan's situation in one important regard, however. Despite the predicted irreversible photochemical destruction of methane, which would deplete atmospheric CH$_4$ and lead to the formation of more complex hydrocarbons within $\sim10~\rm Myrs$ \citep{1984ApJS...55..465Y}, Titan's atmosphere maintains a $1-5\%$ methane composition \citep{2005Natur.438..779N}. While the source of methane in Titan's current atmosphere is still a mystery \citep[e.g.,][]{2006P&SS...54.1177A}, Huygens Probe's measurements of primordial noble gases indicate that Titan's current atmospheric methane is likely a consequence of outgassing from its interior \citep{2005Natur.438..779N,2010JGRE..11512006N}.

Note that all these results are predicated on the surface being unreactive, being neither a source nor sink of atmospheric gases. However, geological or biological processes could invalidate this assumption and alter the model prediction. In the 1 bar surface scenarios presented here, we implicitly assume the presence of a large surface reservoir of liquid. Long-term interactions between hydrocarbons and surface liquids via aqueous chemistry could play an important role and can be explored in future works. In the 1000 bar surface scenarios, the ``surface" temperature far exceeds the solidus of silicates (see Figure~\ref{fig:PTprofile}), suggesting that the ``surface" may be in a magma ocean regime where volatile-rock interactions dictate the lower boundary conditions, as explored by \citet{shorttle2024distinguishing}.

Moreover, a limitation of this study is that we did not account for the feedback effects of chemistry and water condensation on the resultant temperature-pressure profile. While the full integration of chemistry into the radiative transfer calculations remains a long-term objective, several recent studies have explored the impact of water condensation on the temperature profile within the convective zone of H$_2$-dominated atmospheres \citep{guillot1995condensation,markham2022convective,ge2023heat,innes2023runaway,2024arXiv240106608L}. Given that water has a large latent heat and a significantly higher molecular weight than the H$_2$-He mixture, the condensation of water could largely alter the mean molecular weight and buoyancy of the atmosphere. Consequently, the region where water clouds form might suppress convection, leading to a superadiabatic temperature gradient in the convective zone. This would, in turn, reduce the vertical transport efficiency of chemical tracers. Given the large uncertainty in the atmospheric opacity and resulting temperature profile, we will defer a detailed investigation of the chemistry feedback to future studies.

\subsection{Heavy Organic Deposit on Shallow Surface with Water Condensation: The Effect of the Choice of Chemical Network}
\label{sec:heavy}
\begin{figure}[h!]
        \centering
        \includegraphics[width = 0.45 \textwidth]{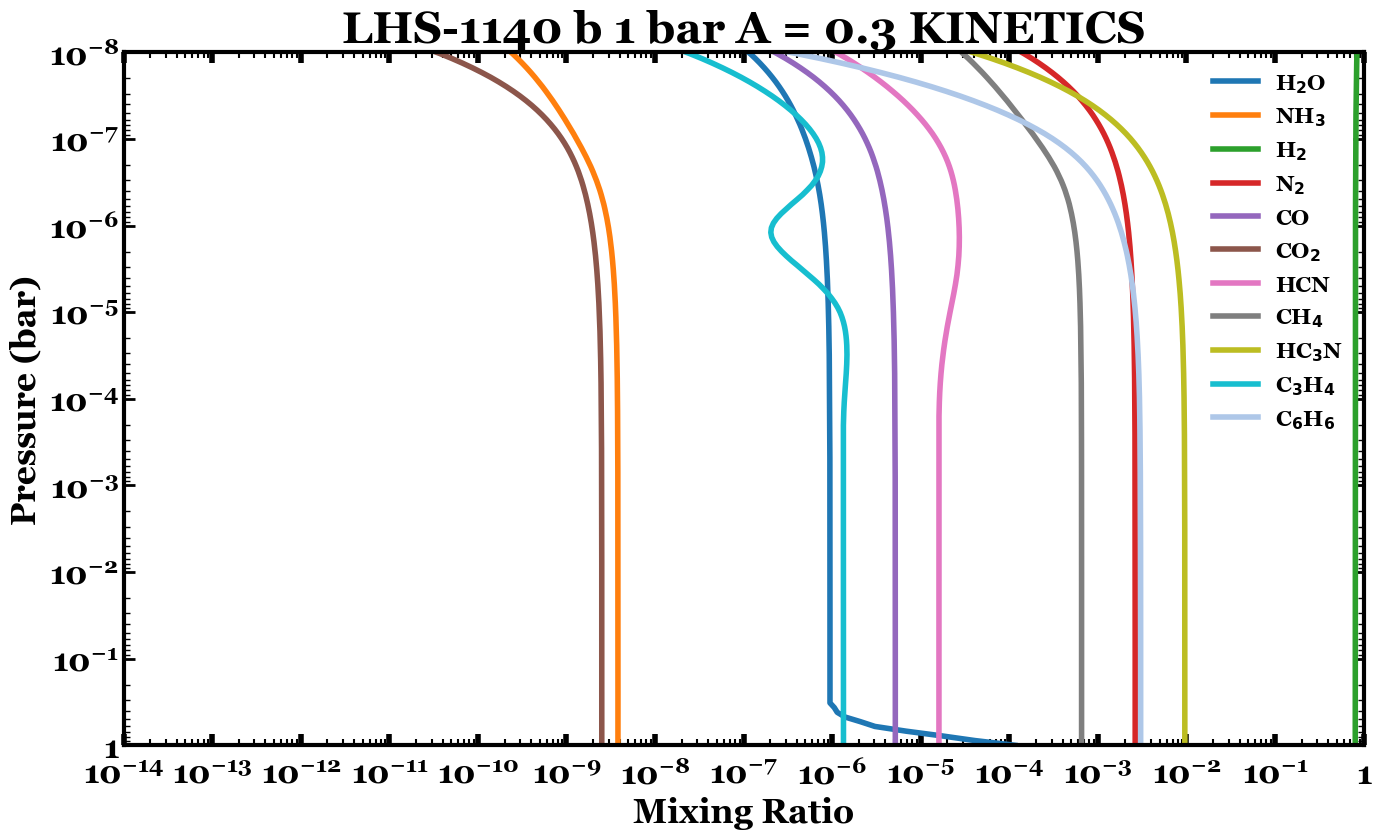}
        \includegraphics[width = 0.45 \textwidth]{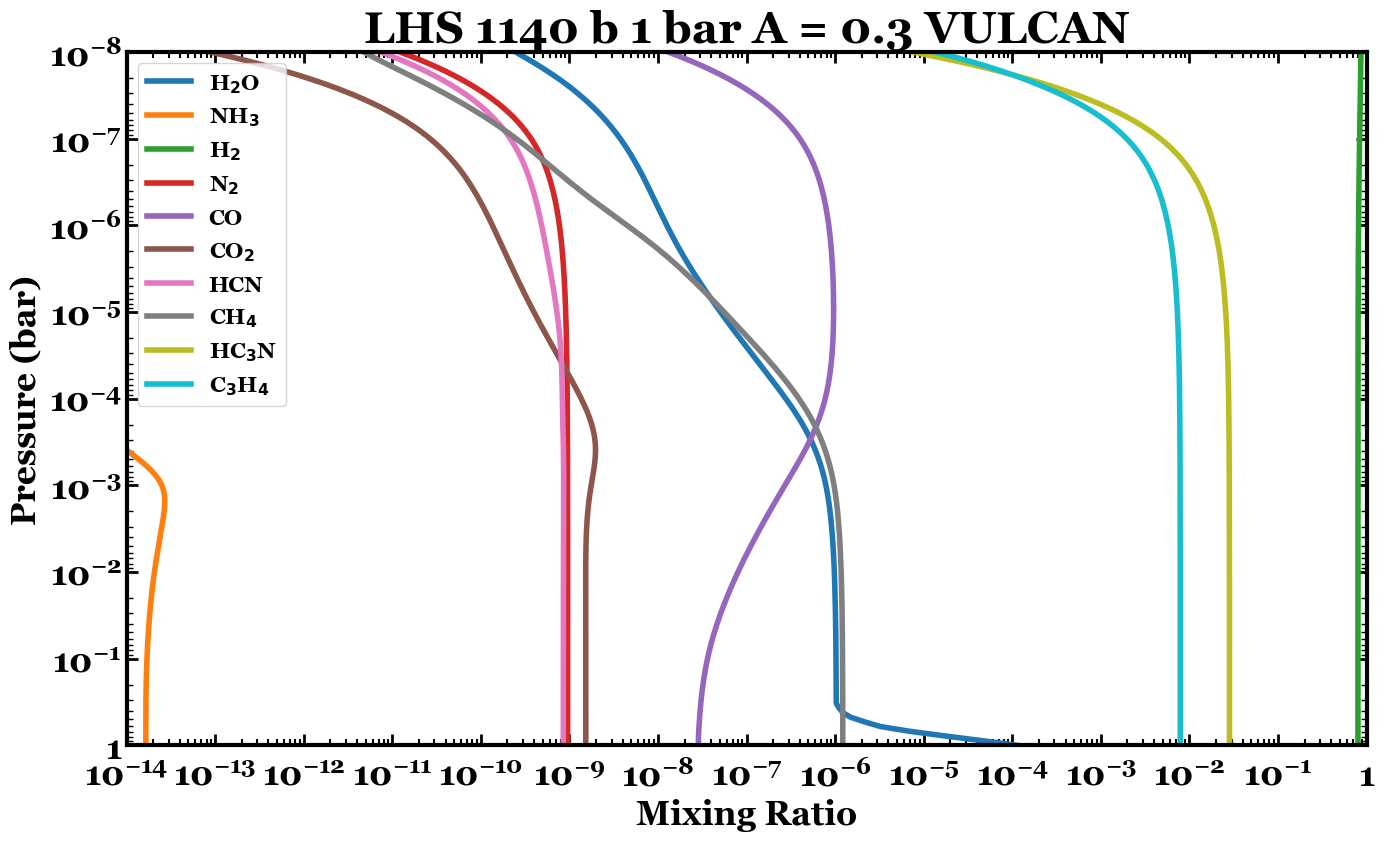} 
        \includegraphics[width = 0.45 \textwidth]{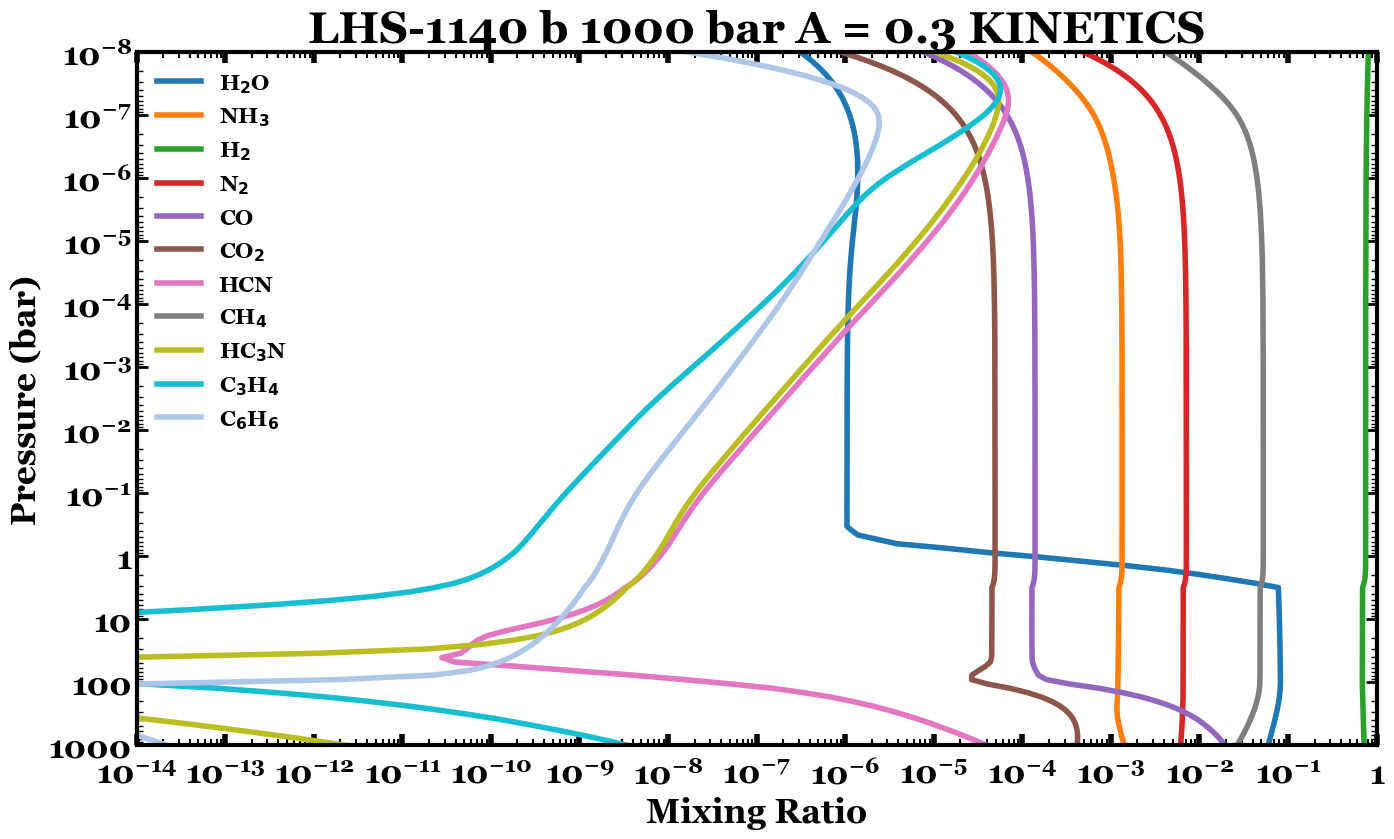} 
        \includegraphics[width = 0.45 \textwidth]{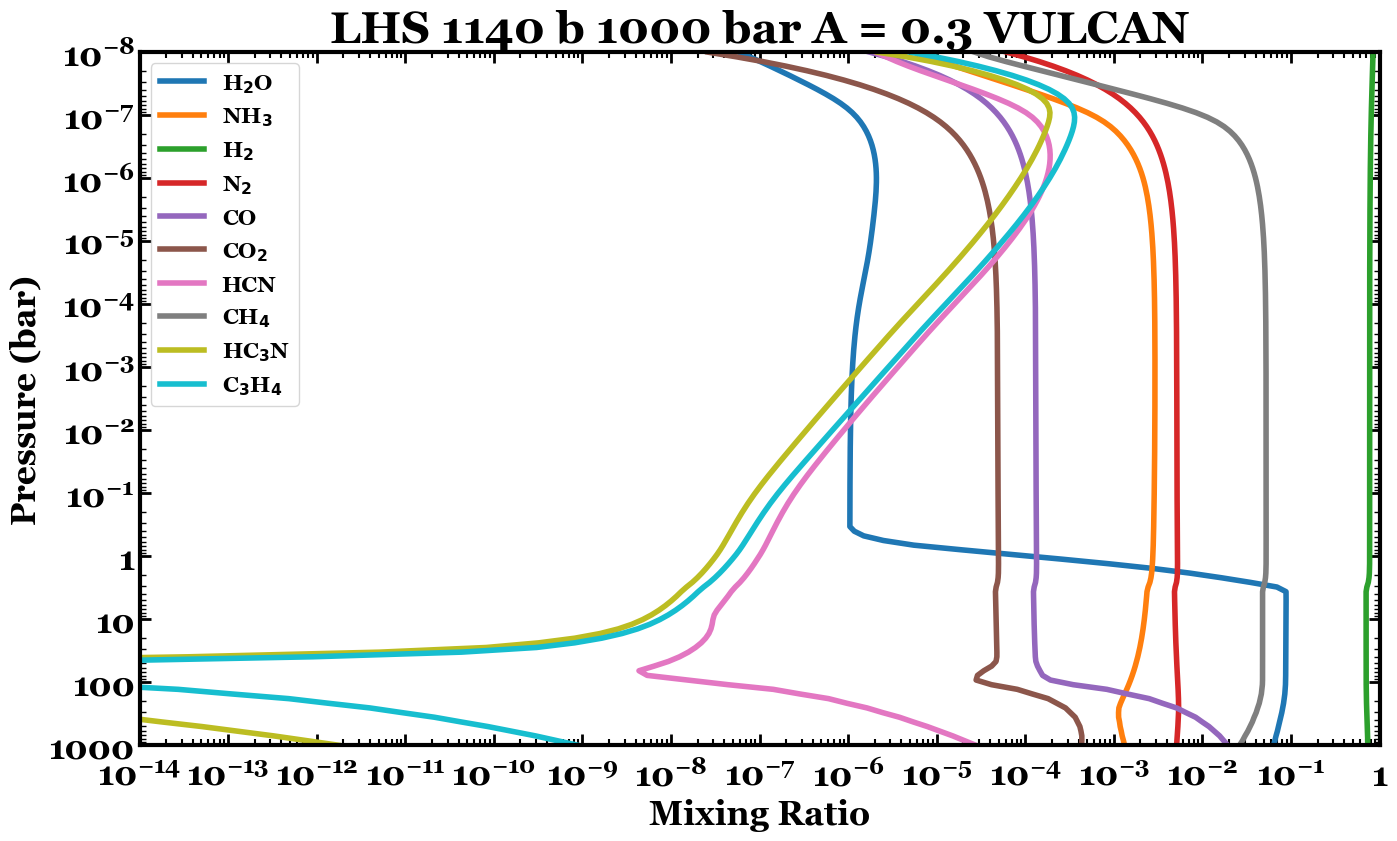} 
    \caption{Simulated VMR profiles for the main chemical species for 1 bar and 1000 bar LHS 1140 b with A = 0.3 using KINETICS and VULCAN. Note that the in KINETICS runs the VMR$_{C_3H_4}$ is the sum of VMR$_{CH_3C_2H}$ and VMR$_{CH_2CCH_2}$.}
\label{fig:compare_vulcan_kinetics}    
\end{figure}

We notice that for the shallow-surface cases with water condensation, the atmosphere ends up being dominated by heavy organics like HC$_3$N and C$_3$H$_4$. While both species are observed in Titan's atmosphere \citep{2023ApJS..266...30Y,2024ESC.....8..406N}, they are not the most abundant carbon and nitrogen carriers. The accumulation of these complex chemical compounds in our simulation is, in part, a result of the limitation of the chemical reaction network used in our model, which lacks reactions to effectively destroy these molecules, omits the production of heavier hydrocarbons and nitriles, and neglects condensation and rain out of such species. This limitation underscores the choice of the chemical network in accurately simulating exoplanet atmospheres. Here, we conduct a comparison study using two chemical networks, VULCAN following \cite{tsai2021inferring} and KINETICS following \cite{moses2016composition}. For the LHS 1140 b 1 bar case with water condensation, both models/networks show the accumulation of HC$_3$N and heavy hydrocarbons as indicated in Figure~\ref{fig:compare_vulcan_kinetics}. Specifically, the two models converge at very similar HC$_3$N mixing ratios. Again, this result is partially caused by the fact that HC$_3$N is the heaviest nitrile considered in both models, HC$_3$N is efficiently recycled by photochemistry under the relevant conditions, condensation of HC$_3$N is not included in the models (and it is abundant enough to condense at the temperatures considered for these planets), and the production of even heavier nitriles is neglected.

The two models differ, however, in the predictions for heavier hydrocarbons. For VULCAN, the most abundant carbon carrier is C$_3$H$_4$, due to insufficient loss processes for this end-product molecule, while for KINETICS, the most abundant carbon carriers are CH$_4$ and C$_6$H$_6$, the latter which is the heaviest stable hydrocarbon considered in the model. Other main nitrogen carriers, such as N$_2$ and HCN are also different between VULCAN and KINETICS. Specifically, VULCAN predicts N$_2$ being the second most abundant nitrogen-carrier, while for KINETICS, HCN is the second most abundant nitrogen-carrier. Even though both models converge at very similar HC$_3$N mixing ratios, the major reactions leading to HC$_3$N in KINETICS are different compared to VULCAN, as CH$_4$ is a lot more abundant in KINETICS than VULCAN. The dominant column-integrated pathway in the stratosphere with KINETICS is the following:

\[\begin{aligned} & \mathrm{HCN}+h \nu \rightarrow \mathrm{CN}+\mathrm{H}, \\ & \mathrm{CH}_4+h \nu \rightarrow \mathrm{CH}+\mathrm{H}_2+\mathrm{H}, \\ & \mathrm{CH}+\mathrm{CH}_4 \rightarrow \mathrm{C}_2 \mathrm{H}_4+\mathrm{H}, \\ & \mathrm{C}_2 \mathrm{H}_4+h \nu \rightarrow \mathrm{C}_2 \mathrm{H}_2+\mathrm{H}_2, \\ & \mathrm{C}_2 \mathrm{H}_2+\mathrm{CN} \rightarrow \mathrm{HC}_3 \mathrm{N}+\mathrm{H},\end{aligned}\]

Here, neither model includes condensation/rainout, dissociation, or polymerization of these high-order hydrocarbon and nitrile species. For example, these species would further be dissociated and polymerized to form complex organic hazes, similar to the formation of hazes in Titan's atmosphere \citep[e.g.,][]{2008P&SS...56...67L}. Thus, we generally conclude that for oxygen-poor exoplanet atmospheres as a result of water condensation, much of the original carbon and nitrogen will end up being deposited in heavier hydrocarbon and nitrile photochemical products, but the exact species and chemical complexity are uncertain due to the limitation of the chemical networks employed by existing photochemical models. We have attempted to add additional pathways to destroy C$_3$H$_4$ and HC$_3$N in VULCAN, which would only lead to the buildup of heavier species. Thus, the resulting built-up of C$_3$H$_4$ and HC$_3$N in the shallow-surface case with condensation is not realistic and can only be treated as a proxy for the production of heavy hydrocarbons and nitriles and polyaromatic-nitrogenated hydrocarbons (PANHs). The atmosphere may still host detectable quantities of HCN, HC$_3$N, both are potentially observable by JWST \citep{2021ApJ...921L..28R, 2023AJ....166...39C}, and some other hydrocarbon and nitrile photochemical products, but they are unlikely to be the dominant carbon and nitrogen constituents, and in such cases, the hazes may be so thick that detection of any species would be difficult.

The significant differences between the VULCAN and KINETICS models, especially for major species such as CH$_4$, C$_6$H$_6$, HCN, and N$_2$, highlight the challenges and limitations in current photochemical modeling. Future studies should focus on expanding the chemical reaction networks and including processes such as condensation, rainout, and the production and polymerization of heavier hydrocarbons and nitriles. Additionally, further experimental and observational data are needed to better constrain the reaction rates and pathways for key photochemical processes. Addressing these issues will improve the reliability of photochemical models in predicting the atmospheric compositions of exoplanets.

\subsection{Simulated Transmission Spectra for Surface Identification}
\label{sec:trans}
\begin{figure}[h!]
        \centering
        \includegraphics[width = \textwidth]{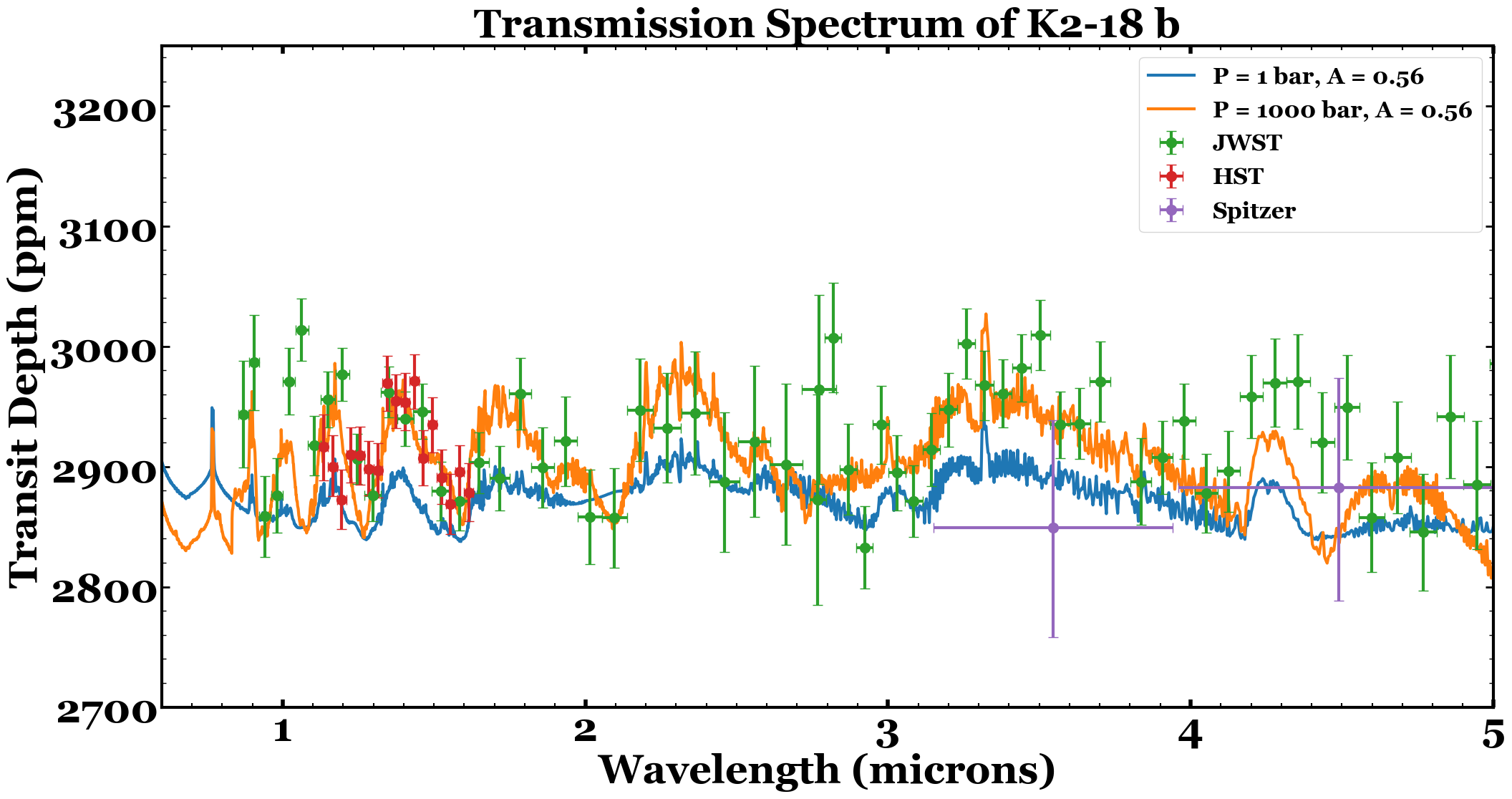} 
        \includegraphics[width = \textwidth]{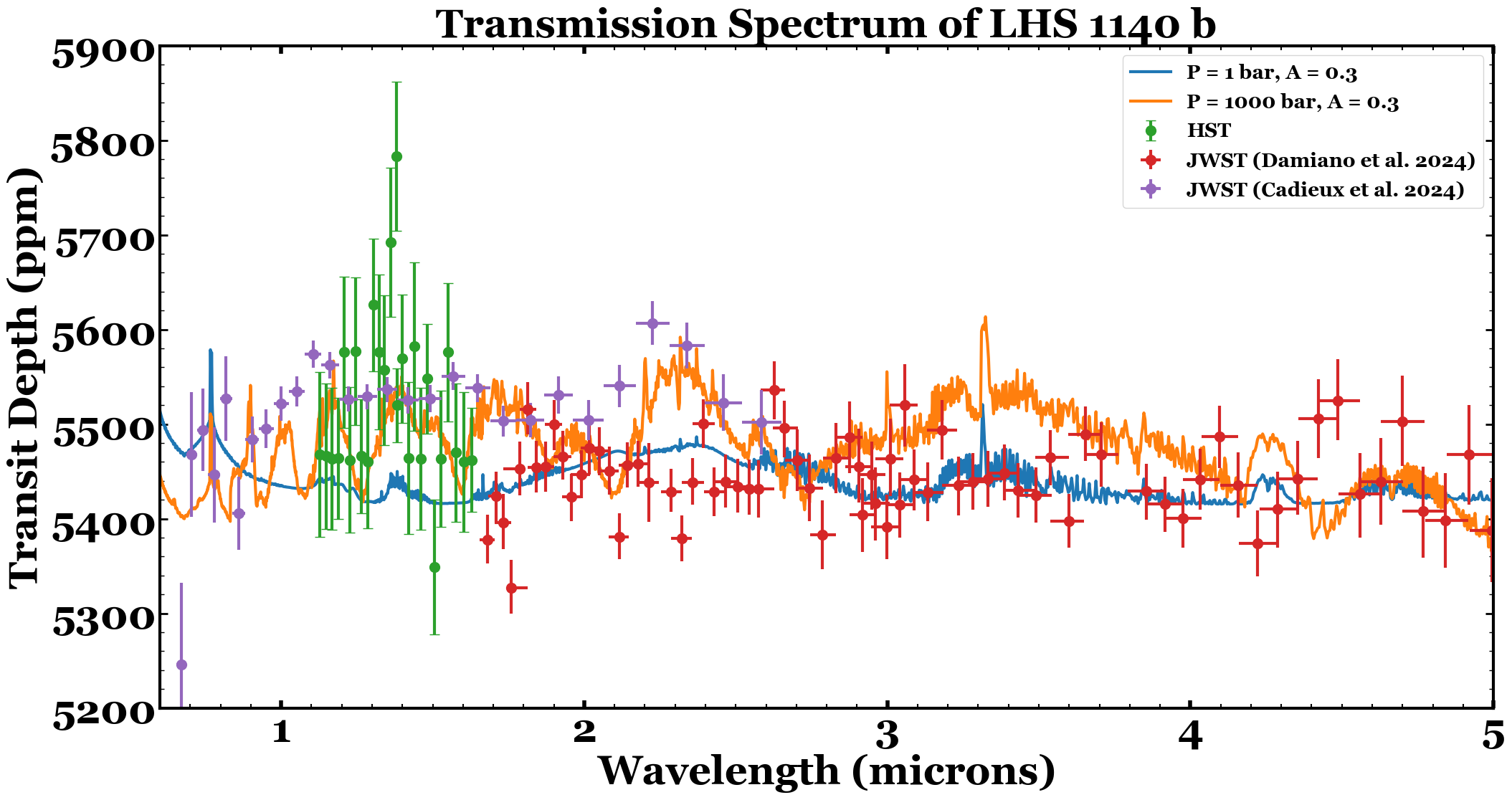}   
    \caption{Simulated transmission spectrum of K2-18 b (top) and LHS 1140 b (bottom) A$\rm_b=0.3$ cases with different surface pressure levels with water condensation. Solid lines are simulation results with the pressure of the bottom surface pressure at 1 bar and 1000 bar. (Top) Green dots are JWST observations reproduced from \citet{madhusudhan2023carbon}. Red dots are HST observations from \citet{benneke2019water}. Purple dots are Spitzer observations from \citet{benneke2017spitzer}. (Bottom) The red and purple dots are JWST observations reproduced from \citet{2024ApJ...968L..22D} and \citet{2024arXiv240615136C, 2024arXiv240312617D}, respectively. Green dots are HST observations from \citet{edwards2020hubble}.}
\label{fig:TransSpec}    
\end{figure}

In light of the recent JWST observations of K2-18 b presented by \citet{madhusudhan2023carbon}, we compare our simulated JWST spectra for K2-18 b with the available reduced data, as shown in Figure~\ref{fig:TransSpec}(top). Observations from HST \citep{benneke2019water} and Spitzer \citep{benneke2017spitzer} are also plotted for a comprehensive comparison. Note that we do not include HC$_3$N and C$_3$H$_4$ in the transmission spectra, as they are merely representative of the high-order hydrocarbon/nitrile species that are generated using our limited chemical network, and the current model does not consider condensation/rainout, dissociation, or polymerization of these high-order species (see discussion in Section~\ref{sec:heavy}). 

Based on the comparison in Figure~\ref{fig:TransSpec}, the low abundance of H$_2$O is indeed consistent with the scenario of water condensation happening below the observable photosphere \citep{madhusudhan2023carbon}, likely as a result of the increased albedo of the planet due to conservatively scattering hazes and water clouds, which decrease its equilibrium temperature (see, however, \citet{2024arXiv240106608L}, who discuss that tropospheric water clouds alone cannot increase the albedo of K2-18b sufficiently to explain the cold tropopause temperatures). Note that the free retrieval and Bayesian-evidence analysis in \cite{madhusudhan2023carbon} shows that K2-18 b is likely depleted in ammonia. Based on previous PANDEXO simulations \citep{hu2021photochemistry, tsai2021inferring}, 3-4 transits could be needed to definitively distinguish the existence of NH$_3$. Indeed, in the observed wavelength range from $0.85-5.17~\mu$m, the absorption features of NH$_3$ are the most prominent around $3~\mu$m, in between bands of CO$_2$ and CH$_4$. Although \cite{madhusudhan2023carbon} infer an upper limit for NH$_3$ from the existing observations, spectra from additional transits would help reduce the error bars in the relevant wavelength region, allowing more definitive constraints or tighter upper limits for NH3. 

The free retrieval of \cite{madhusudhan2023carbon} indicates $\sim1\%$ CH$_4$ and $\sim1\%$ CO$_2$ in the atmosphere of K2-18 b, with $1\sigma$ abundance estimation within one order of magnitude, which disfavors a shallow surface with water condensation. Instead, of our two scenarios, the deep-surface model so far best fits the data (see also \citet{wogan2024jwst}), the CO$_2$ VMR ($10^{-3.52}$) is within $2\sigma$ and the CH$_4$ VMR ($10^{-1.30}$) is within $1\sigma$ error bars of the 1-offset free retrieval. In \cite{madhusudhan2023carbon} and \cite{hu2021photochemistry}, it was proposed that a 1-bar ocean planet could explain the observed 1\% CH$_4$ and 1\% CO$_2$ abundances. However, recent work \citep{wogan2024jwst} disfavors this scenario and instead finds a much lower CH$_4$ mixing ratio (1 ppm) for the ocean case, unless there is a significant flux of CH$_4$ from the interior/ocean. Recent work by \citet{shorttle2024distinguishing} also attempts to explain the depletion of ammonia and CO/CO$_2$ abundance assuming a magma ocean bottom boundary; however, more observations are needed to ascertain the existence and abundance of NH$_3$ and CO to test such a scenario.

We also simulated the transmission spectrum of LHS 1140 b, which has been observed by JWST \citep{2024ApJ...968L..22D, 2024arXiv240312617D,2024arXiv240615136C}, as shown in Figure~\ref{fig:TransSpec} (bottom). Even though the reduced transmission spectra of \citet{2024ApJ...968L..22D} and \citet{2024arXiv240312617D, 2024arXiv240615136C} look quite different in the overlapping region between 1.67 and $2.8~\mu m$, both are rather flat spectra. This likely rules out LHS 1140 b being a mini-Neptune with a deep H$_2$-rich atmosphere \citep{2024ApJ...968L..22D, 2024arXiv240312617D, 2024arXiv240615136C}. The shallow-surface scenario with water condensation is more consistent with the current observations of LHS 1140 b. When LHS 1140 b has albedo equal to or above 0.3, it is cold enough so that the saturation vapor mixing ratio at the surface is significantly lower than the 100$\times$ solar metallicity equilibrium mixing ratio of water, leading to enhanced C/O ratio in the atmospheric region above the tropopause cold trap. The resulting planet is depleted in NH$_3$, CH$_4$, CO, and CO$_2$ (see the blue transmission spectrum in Figure~\ref{fig:TransSpec} (bottom)). Without the efficient methane and ammonia recycling for the shallow-surface case, complex hydrocarbons and nitriles would be produced (in our model represented by C$_3$H$_4$ and HC$_3$N due to the limited chemical network), leading to abundant photochemical hazes to flatten the spectrum, though our current model has not included the potential haze impacts. %Note that the KINETICS model produces more CH$_4$ (\td{red} transmission spectra) than the VULCAN model (\td{blue} transmission spectra), as its carbon carrier ends at CH$_4$ and C$_6$H$_6$, instead of C$_3$H$_4$, due to the choice of different chemical networks (see discussion in Section~\ref{sec:heavy}). However, both models predict depletion in CO/CO$_2$, which matches the spectra pretty well.

Note that similar effects would be achieved if LHS 1140 b were to have an intrinsically high atmospheric C/O ratio (so CO$_2$ would not be the main carbon carrier), water condensation, and a shallow surface. In addition to the high-mean-molculear-weight atmosphere suggested by \citet{2024ApJ...968L..22D} and \citet{2024arXiv240615136C}, the JWST observations of LHS 1140 b also could be consistent with a very hazy, shallow, H$_2$-dominated atmosphere with an unreactive surface and water condensation along with the enhanced albedo of the planets due to clouds/hazes or a high atmospheric C/O ratio.
 
\section{Summary}

The effect of condensation of water is critical for the atmospheric evolution of cool exoplanets. By photochemical modeling of the atmosphere with water condensation included, we identify the following differences compared to the non-condensation scenario:
\begin{itemize}
    \item Water condensation is shown to significantly impact the atmospheric chemistry of cool-to-temperate sub-Neptunes, particularly those with shallow surfaces, by increasing the C/O ratio in the photochemically active region of the atmosphere and altering the main carbon and nitrogen carrier species, leading to the depletion of oxygen-carrying carbon species including CO$_2$, CO, and NH$_3$ and the enhancement of heavier-order hydrocarbons and nitriles.
    \item In scenarios where sub-Neptunes have cool, shallow surfaces, and when the saturation vapor pressure falls below the partial pressure of water, the atmospheric carbon evolves towards heavier hydrocarbons and nitriles instead of CO$_2$ or CO, leading to a significant reduction in oxygen-bearing molecules. Such planets may be expected to have very hazy atmospheres. In addition to a possible high mean-molecular-weight atmosphere for LHS 1140 b, as has been suggested by \citet{2024ApJ...968L..22D, 2024arXiv240615136C}, a thin, H$_2$-dominated atmosphere with water condensation, a high resulting stratospheric C/O ratio, and abundant high-altitude photochemical hazes could offer a possible explanation for the observed JWST observations of LHS 1140 b, explaining the lack of detected spectral features of NH$_3$, CH$_4$, CO$_2$, and CO.
    \item Our simulated transmission spectra that account for varying conditions of water condensation and atmospheric surface pressure reveal distinct spectral signatures that can be instrumental in characterizing sub-Neptunes. The JWST observed transmission spectra of K2-18 b align more closely with the deep-surface mini-Neptune case, agreeing \citet{wogan2024jwst}, although CO$_2$ is still underpredicted by such models. For LHS 1140 b, we propose an alternative solution for its observed relatively flat spectra; instead of having a high-molecular-weight atmosphere, the observed flat spectra are also consistent with a super-Earth-like exoplanet, with significant water condensation.
\end{itemize}

\section{Acknowledgements}
X. Yu and Z. Huang are supported by the Heising-Simons Foundation grant 2023-3936. X. Yu is also supported by the NASA Planetary Science Early Career Award 80NSSC23K1108. X. Yu and J. Krissansen-Totton are supported by the NASA Habitable Worlds Program Grant 80NSSC24K0075. S.-M. Tsai is supported by NASA through Exobiology Grant No. 80NSSC20K1437 and Interdisciplinary Consortia for Astrobiology Research (ICAR) Grant Nos. 80NSSC23K1399, 80NSSC21K0905, and 80NSSC23K1398. J. Moses is supported by NASA Exoplanet Research Program grant 80NSSC23K0281. K. Ohno is supported by JSPS Overseas Research Fellowship and JSPS KAKENHI Grant Number JP23K19072. X. Zhang is supported by the National Science Foundation grant AST2307463 and NASA Exoplanet Research grant 80NSSC22K0236. J. Fortney and X. Zhang acknowledge support from the NASA Interdisciplinary Consortia for Astrobiology Research (ICAR) grant 80NSSC21K0597. This work also benefited from the 2022 and 2023 Exoplanet Summer Programs in the Other Worlds Laboratory (OWL) at the University of California, Santa Cruz, a program funded by the Heising-Simons Foundation.

\section{Appendix}

\begin{figure}
    \centering
        \includegraphics[width = \textwidth]{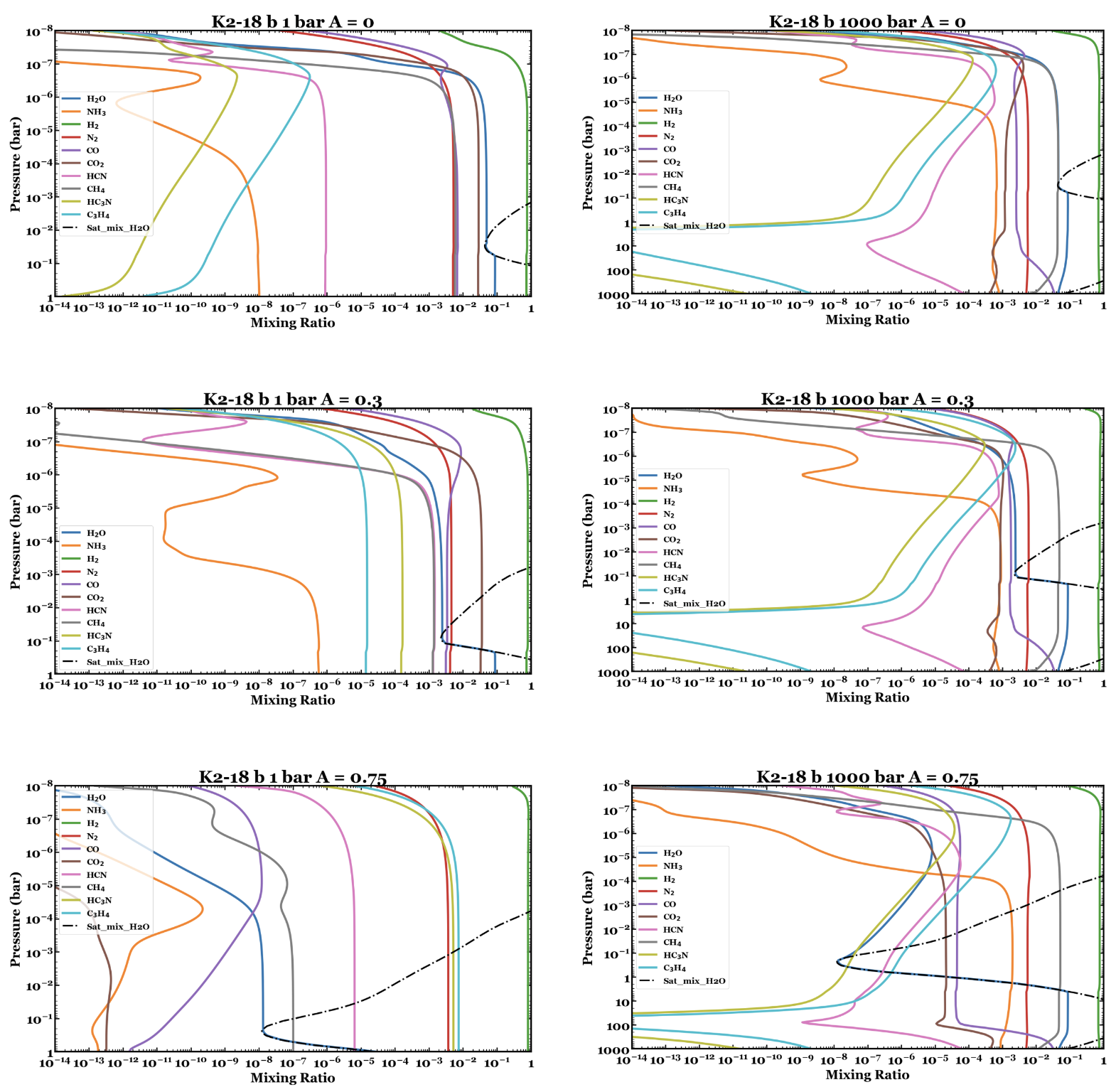}  
    \caption{Simulated VMR profiles for the main chemical species for K2-18 b with water condensation for the albedo cases that are not presented in the main text. The left and right columns are for the shallow and deep surface cases, respectively. (top) A$\rm_b=0$ (T$_{\rm eq} = 278~\rm K$), (middle) A$\rm_b=0.3$ (T$_{\rm eq} = 254~\rm K$), (bottom) A$\rm_b=0.75$ (T$_{\rm eq} = 197~\rm K$).}
\label{fig:K2_appendix}    
\end{figure}

\begin{figure}
    \centering
        \includegraphics[width = \textwidth]{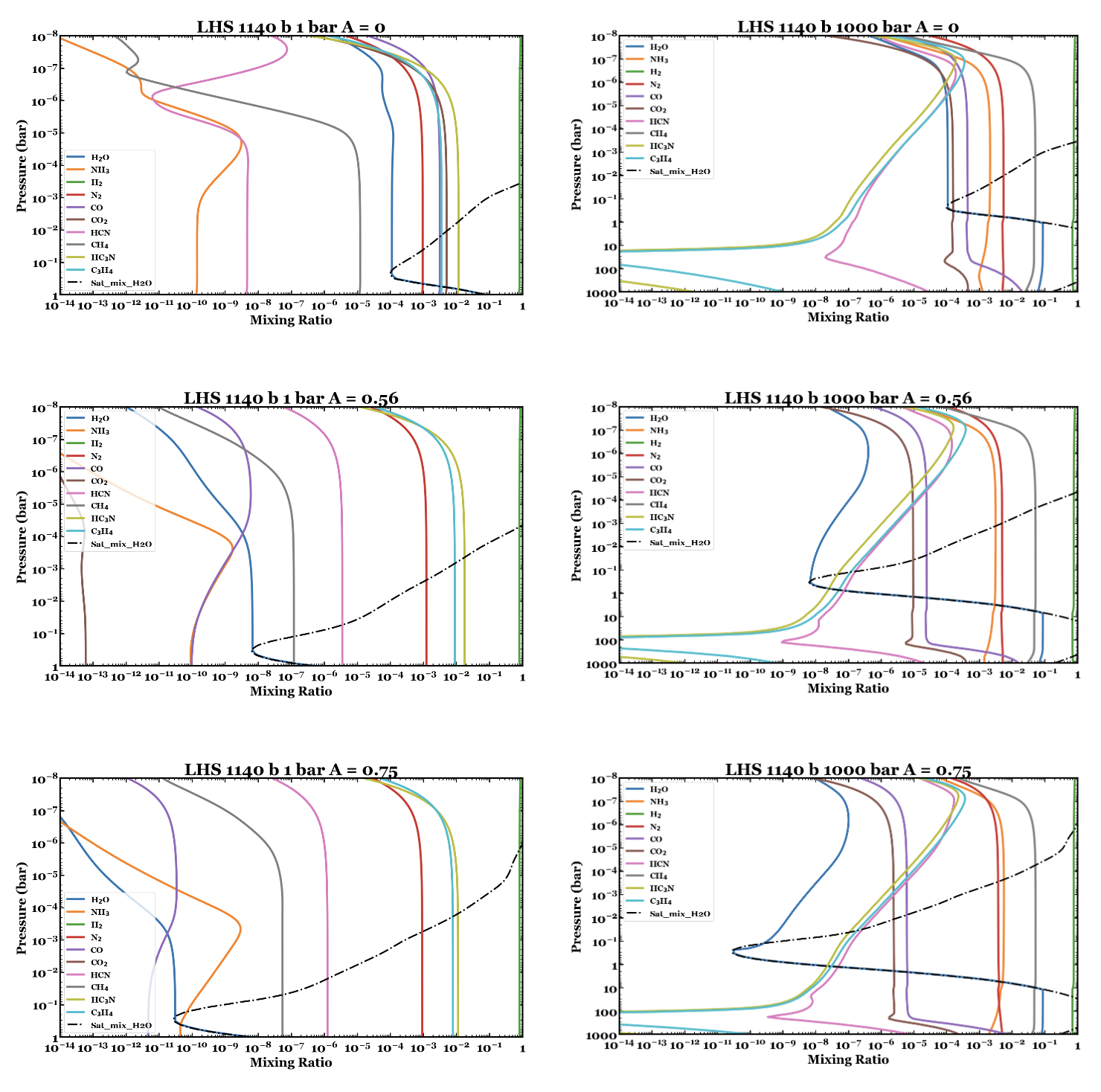}  
    \caption{Simulated VMR profiles for the main chemical species for LHS 1140 b with water condensation for the albedo cases that are not presented in the main text. The left and right columns are for the shallow and deep surface cases, respectively. (top) A$\rm_b=0$ (T$_{\rm eq} = 234~\rm K$), (middle) A$\rm_b=0.56$ (T$_{\rm eq} = 191~\rm K$), (bottom) A$\rm_b=0.75$ (T$_{\rm eq} = 165~\rm K$).}
\label{fig:LHS_appendix}    
\end{figure}

The steady-state atmospheric compositions for the albedo cases that are not presented in the main text for K2-18 b are shown in Figure~\ref{fig:K2_appendix} and for LHS 1140 b in Figure~\ref{fig:LHS_appendix}.

\bibliography{references}{}
\bibliographystyle{aasjournal}

\end{document}